\documentstyle[aps,epsfig,float,amstex]{revtex}
\newcommand{\be}{\begin{equation}}
\newcommand{\ee}{\end{equation}}
\newcommand{\beq}{\begin{eqnarray}}
\newcommand{\eeq}{\end{eqnarray}}
 
\begin{document}
    
\def\gC{\mbox{\boldmath $C$}}
\def\gZ{\mbox{\boldmath $Z$}}
\def\gR{\mbox{\boldmath $R$}}
\def\gN{\mbox{\boldmath $N$}}
\def\ua{\uparrow}
\def\da{\downarrow}
\def\a{\alpha}
\def\b{\beta}
\def\g{\gamma}
\def\G{\Gamma}
\def\d{\delta}
\def\D{\Delta}
\def\e{\epsilon}
\def\ve{\varepsilon}
\def\z{\zeta}
\def\h{\eta}
\def\th{\theta}
\def\k{\kappa}
\def\l{\lambda}
\def\L{\Lambda}
\def\m{\mu}
\def\n{\nu}
\def\x{\xi}
\def\X{\Xi}
\def\p{\pi}
\def\P{\Pi}
\def\r{\rho}
\def\s{\sigma}
\def\S{\Sigma}
\def\t{\tau}
\def\f{\phi}
\def\vf{\varphi}
\def\F{\Phi}
\def\c{\chi}
\def\w{\omega}
\def\W{\Omega}
\def\Q{\Psi}
\def\q{\psi}
\def\de{\partial}
\def\inf{\infty}
\def\ra{\rightarrow}
\def\bra{\langle}
\def\ket{\rangle}
\title{On-Site Repulsion as the Source of  Pairing in Carbon
Nanotubes and  Intercalated Graphite}
\author{Enrico Perfetto, Gianluca Stefanucci and Michele Cini}
\address{Istituto Nazionale per la Fisica della Materia, 
Dipartimento di Fisica,\\
Universita' di Roma Tor Vergata, 
Via della Ricerca Scientifica, 1-00133\\
Roma, Italy}
\maketitle

\begin{abstract} 
We show that different non-conventional superconductors have 
one fundamental feature in common: pair eigenstates of the 
Hamiltonian are repulsion-free, the $W=0$ pairs. In 
extended  Hubbard models,  pairing can occur for resonable  
parameter values. For $(N,N)$ nanotubes the binding 
energy of the pair  depends strongly on the filling and decreases 
towards a reduced but nonzero value for the graphite sheet $N \rightarrow \infty$.
\end{abstract}

\vspace{1cm}

{\small

There is experimental evidence that the critical Temperature $T_{c}$ 
in alkali-graphite intercalation compounds (GIC) C$_{x}$M (where M is 
a given alkali metal) grows as $x$ decreases\cite{belash}. 
Under high-pressure, high metal concentration samples such as 
C$_{6}$K, C$_{3}$K, C$_{4}$Na, C$_{3}$Na, C$_{2}$Na, C$_{2}$Li have 
been synthesized; for C$_{2}$Na the value of $T_{c}$ is 5 K while for 
C$_{2}$Li, $T_{c}$=1.9 K; quite recently Potassium\cite{bockrath2} and Lithium\cite{gao} 
have been intercalated also in single- and multi-wall carbon 
nanotubes\cite{Iijima} 
up to high concentration (the highest metal concentration was obtained with 
Lithium in C$_{2}$Li) and a net charge transfer was observed between the alkali-metals 
and the carbon atoms. The alkali-metals  cause little structural 
deformation, but increase the filling of  the 
original bands. Nanotubes close to half filling  are deemed to be 
 Luttinger liquids down to milli-Kelvin  temperatures\cite{balents}\cite{sol}.
In this letter we use the Hubbard Hamiltonian $H$ on the honeycomb lattice 
to represent the valence bands of carbon single-wall $(N,N)$ nanotubes 
and propose a symmetry-driven configuration interaction pairing mechanism 
which works away 
from half-filling. 
We present  analytic expressions  for the effective interaction and obtain 
the binding energy for $(N,N)$ armchair nanotubes; in the case $N=1$ we 
verify these analytic results by exact diagonalization and get high-precision agreement. 
Starting from the undoped system we find that the pair binding energy grows 
as the number 
of electrons per C atom increases. Furthermore, we obtain stronger binding 
in nanotubes than in graphite sheets and this suggests a higher critical 
temperature for the former. This is also supported by the 
measurements of a $T_{c}\approx 15$ K in the 4 Angstrom single-wall 
nanotube (SWNT) by Tang {\em et al.}\cite{tang}.

Using  standard notation, the full Hamiltonian reads 
\begin{equation}
H=H_{0}+ W= t\sum \limits_{ \langle {\bf r},{\bf r}' \rangle }\sum
 \limits_{\sigma}\left(c^{\dagger}_{{\bf r},\sigma}c_{{\bf r}',\sigma}+h.c. \right)+
 U\sum \limits_{{\bf r}} \hat{n}_{{\bf r},\uparrow }
\hat{n}_{{\bf r},\downarrow  } ,
\label{hub}
\end{equation}
where ${\bf r}$ denotes the honeycomb site, the sum runs over the
pairs $\langle {\bf r},{\bf r}'\rangle$ of nearest neighbour carbon atoms 
and $t$ is the hopping parameter. The one-body eigenvalues $\ve^{\pm}({\bf k})$, 
(- for the bonding and + for the antibonding bands) are readily obtained, and 
the Fermi line has $C_{2v}$ symmetry for the nanotubes ($C_{6v}$ for 
the graphite sheet). Here, we assume that the Fermi 
level $\ve_{F}$ lies in the + band. The Hamiltonian in Eq.(\ref{hub}) 
admits {\em two-body} singlet eigenstates with no
double occupancy and we shall refer to them 
as $W=0$ pairs. The particles forming a $W=0$ pair have no 
direct interaction and are the main candidates to achieve bound 
states in purely repulsive Hubbard models already used for the 
Cuprates\cite{cibal}\cite{EPJB1999}\cite{EPJB2001}.
We note that such states are also building bricks of the 
ground state of Hubbard and related models at half 
filling\cite{ssc2001}\cite{jop2002}\cite{tobepub}.

Recently we obtained\cite{EPJB2001}\cite{IJMPB2000} a general criterion to get 
{\em all} the $W=0$ pairs. We can do that in terms of the Optimal Group 
${\cal G}$ of the Hamiltonian, that we define as a symmetry Group that 
justifies the degeneracy of the single particle energy levels. We may 
say that an irreducible representation (irrep) $\eta$ is {\em represented} in the 
one-body spectrum of $H$ if at least one of the one-body levels belongs to $\eta$. 
Let ${\cal E}$ be the set of the irreps of ${\cal G}$ which are 
represented in the one-body spectrum of $H$ (${\cal E}$ includes all 
the irreps when ${\cal G}$ is Abelian). Let $|\q\rangle$ be a two-body eigenstate of 
the kinetic energy $ H_{0}$ with vanishing $z$ component of the spin. Then, it holds the 
{\it W=0 Theorem}:  
\begin{equation}
\eta \notin  {\cal E} \Leftrightarrow W P^{(\eta)}|\q\rangle=0 \;
\label{theo}
\end{equation}
where $P^{(\eta)}$ is the projection operator on the irrep $\eta$.
In other terms, any nonvanishing projection 
of $|\q\rangle$ on an irrep
{\em not} contained in ${\cal E}$, is an eigenstate of $ H_{0}$ 
with no double occupancy. The singlet component of this state is a  
$W=0$ pair. Conversely, any pair belonging to an irrep represented 
in the one-body
spectrum must have positive $W$ expectation value. If a subgroup of 
${\cal G}$ is available, the $\Rightarrow$ implication still holds.
Using the Space Group, we find that in the vanishing quasi-momentum  sector 
the only $W=0$ pairs belong to the pseudoscalar irrep $A_{2}$. 
Let $(a,b)$ denote the basis of the Bravais lattice and 
$u\left({\bf k},\z\right)$ the periodic part of 
the Bloch function of quasi-momentum ${\bf k}$, with $\z =a,b$. The pair 
wavefunction reads\cite{psc}
\begin{eqnarray}
\q^{[A_{2}]}_{\z_{1},\z_{2}}\left({\bf k}, {\bf R}_{1} , {\bf R}_{2} 
\right) &=& 
\sin \left( k_{x}(X_{1}-X_{2})\right) \times 
\nonumber \\ &\times& \frac{1}{\sqrt{2}} 
\left[u^{\ast}\left({\bf k},\z_{1}\right)
u^{\ast}\left(-{\bf k},\z_{2}\right) 
e^{ik_{y}(Y_{1}-Y_{2})}-
u^{\ast}\left({\bf k},\z_{2}\right) 
u^{\ast}\left(-{\bf k},\z_{1}\right)
e^{-ik_{y}(Y_{1}-Y_{2})}\right]\, \chi_{0}, 
\end{eqnarray}
with ${\bf R}_{i}=(X_{i},Y_{i})$ the origin of the 
cell where the particle $i$ lies. We can verify by direct 
inspection that $\q^{[A_{2}]}_{\z_{1},\z_{2}}({\bf k},{\bf R}_{1},{\bf R}_{2})$ 
vanishes for $X_{1}=X_{2}$, that is the two-body singlet 
wavefunction vanishes if the particles lie on the same annulus of 
the $(N,N)$ tube. As a consequence 
$\q^{[A_{2}]}_{\z_{1},\z_{2}}({\bf k},{\bf R}_{1},{\bf R}_{2})$ is 
an eigenstate of the kinetic energy $H_{0}$ [with eigenvalue 
$2\ve({\bf k})$] and of the on-site Hubbard repulsion $W$ with 
vanishing eigenvalue of the latter, that is 
$\q^{[A_{2}]}_{\z_{1},\z_{2}}({\bf k},{\bf R}_{1},{\bf R}_{2})$ is a 
$W=0$ pair. Remarkably, $\q^{[A_{2}]}=0$ when the {\em transverse} component 
$k_{y}=0$.  

The effective interaction $W_{\rm eff}$ between the particles of a 
$W=0$ pair can be obtained  analytically by a canonical
transformation in the spirit of Ref.\cite{EPJB1999}. Letting 
$n^{(0)}_{\nu}({\bf p})$ denote the non-interacting occupation number 
in band $\nu$ with wavevector ${\bf p}$, we find 
\begin{eqnarray}
W_{\rm eff}({\bf k},{\bf k^{\prime}},E)&=& 
2\sum_{\hat{O} \in C_{2v}} \chi ^{(A_{2})} (\hat{O}) 
\sum_{{\bf p},\n}\;[1-n^{(0)}_{+}(\hat{O}{\bf k}' +{\bf k} 
+{\bf p})] \; n^{(0)}_{\nu}({\bf p})\times\nonumber\\ 
&\times& \frac{U_{\n}(\hat{O}{\bf k}' +{\bf k} +{\bf p},
-{\bf k},\hat{O}{\bf k}',{\bf p}) \, 
U_{\n}({\bf k},{\bf p},\hat{O}{\bf k}'+{\bf k}+{\bf p},-\hat{O}{\bf k}')}
{\varepsilon^{+}(\hat{O}{\bf k}'+{\bf k}+{\bf p})-\varepsilon^{\n}({\bf p})+
\varepsilon^{+}({\bf k}')+\varepsilon^{+}({\bf k})-E}
\label{scattop}
\end{eqnarray}
where  $ \chi ^{ (\eta) }(\hat{O})$ is the character in $\eta$ of the  operation 
$\hat{O}$ of $C_{2v}$, $E$ is the interacting pair energy and 
$U_{\n}({\bf k}_{1},{\bf k}_{2},{\bf k}_{3},{\bf k}_{4})$ 
is the interaction vertex, with incoming legs ${\bf k}_{3}$ and ${\bf k}_{4}$  
in band + and outgoing  ${\bf k}_{1}$ in band + and ${\bf k}_{2}$ 
legs in band $\n$. The effective Schr\"odinger equation for the 
pair reads
\begin{equation}
\left[2\varepsilon({\bf k}) +W_{F}+F({\bf k},E)\right]
\,a_{\bf k}+\sum_{{\bf k}'\in {\cal D}/4}\,W_{\rm eff}\,({\bf k},{\bf 
k}',E)\,
a_{{\bf k}'} =E a_{\bf k}\;,
\label{cooplike}
\end{equation}
where $W_{F}$ is the first-order self-energy shift and 
\begin{equation}
F({\bf k},E) =
- 2\d({\bf k}-{\bf k}') \sum_{{\bf p},\n} \sum_{{\bf q}}
\frac{[1-n^{(0)}_{+}({\bf k} +{\bf p} -{\bf q})]\;
 [1-n^{(0)}_{+}({\bf q})]\;n^{(0)}_{\nu}({\bf p})}
{\varepsilon^{+}({\bf k}+{\bf p}+{\bf q})-\varepsilon^{\n}({\bf p})+
\varepsilon^{+}({\bf q})+\varepsilon^{+}({\bf k})-E}\;
|U_{\n}({\bf k},{\bf p},{\bf k}+{\bf p}-{\bf q},{\bf q}) |^{\,2}
\end{equation}
is the {\em forward scattering} term which does not contains any 
direct interaction between the particles of the pair. 
Eq.(\ref{cooplike}) requires a self-consistent calculation of $E$ 
(since $W_{\rm eff}$ and $F$ are $E$-dependent). The indices ${\bf k}$ and ${\bf k}'$ run 
over $1/4$ of the empty part of the  FBZ and we denoted such a set of 
wavevectors as ${\cal D}/4$. We show below that 
$E=2\ve_{F}+W_{F}+F_{\rm min}({\bf k}_{F})+\D$, with a positive binding 
energy $-\D$ of the $W=0$ pair; here $F_{\rm min}({\bf k}_{F})$ is the 
minimum value of $F({\bf k},E)$ among the ${\bf k}_{F}$-wavevectors on the Fermi 
line. 
\begin{figure}[H]
\begin{center}
	\epsfig{figure=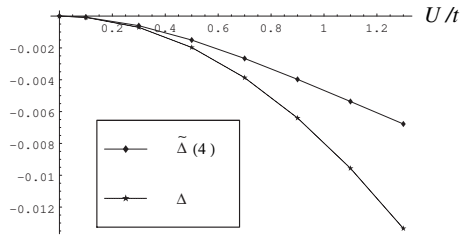,width=6cm}\caption{\footnotesize{
	Comparison between $\D$ and $\tilde{\D}(4)$ in units of $t$ versus $U/t$. 
	}}
\label{comparison}
\end{center}
\end{figure}

We got a direct verification that pairing actually occours  by exact 
diagonalization for the $(1,1)$ nanotube of length $L=2$ (in units of the lattice 
spacing) and periodic boundary conditions.
We define, following Refs.\cite{cibal}\cite{bal},
\begin{equation}
{\tilde \Delta}({\cal N}+2)=E({\cal N}+2)+E({\cal N})-2E({\cal N}+1), 
\label{delta}
\end{equation}
where $E({\cal N})$ is the ground state energy with ${\cal N}$ electrons (referenced 
to the electron vacuum); $|{\tilde \Delta}({\cal N}+2)|$ is one definition of the pairing 
energy. In previous studies of $W=0$ pairing in finite systems we 
found\cite{cibal}\cite{EPJB2001}\cite{cbs3} that at least at weak coupling  
${\tilde \Delta}$ agrees well with $\D$ as obtained by the 
canonical transformation. For the $(1,1)$ nanotube with ${\cal N}=2$, we can see 
in Fig.(\ref{comparison}) that the agreement between $\tilde{\D}(4)$ and $\D$ 
is again very good up to $U/t\approx 1$. However, we emphasize that 
$\tilde{\D}(4)$ decreases up to a characteristic value of $U/t \sim 4 \div 5$,
where a minimum is reached\cite{psc}; at the minimum ${\tilde \Delta}(4) 
\sim -0.018\; t$. The ratio of the second derivatives with respect to $U/t$ at 
$U=0$ was estimated by using best fits and turned out to be $1.00003$, while 
the first derivative vanishes. The binding energy for $U \lesssim t$ 
is in the 10$^{-3}t$ range and by analysing the four-body ground-state wavefunction we 
assessed that the symmetry is $^{1}E_{1}$ as predicted by the canonical 
transformation. Here, $E_{1}$ is a twice degenerate irrep of the Optimal 
Group\cite{psc}  ${\cal G}$ that breaks into $A_{1}\oplus B_{1}$ in $C_{2v}$.  
This result encourages us to proceed with larger and more 
physical systems. 

\begin{center} 
\begin{tabular}{|l||c|c|c|}
\hline
$L$ & $N=2$  & $N=4$ & $N=6$ \\
\hline
\hline
\hline
10 & 
\begin{tabular}{|r|llll|}
\hline
$\ve_{F}$ & 0.8 & 0.9 & 1.0 & 1.1 \\
\hline
$-\Delta$ & 86 & 86 & 82 & 84 \\
\hline 
$-V$ & 2.5 & 2.5 & 2.4 & 2.8 \\
\hline 
\end{tabular}
& 
\begin{tabular}{|r|llll|}
\hline
$\ve_{F}$ & 0.8 & 0.9 & 1.0 & 1.1 \\
\hline
$-\Delta$ & 44 & 47 & 47 & 37 \\
\hline 
$-V$ & 2.2 & 1.5 & 1.5 & 1.7 \\
\hline 
\end{tabular}
&
\begin{tabular}{|r|llll|}
\hline
$\ve_{F}$ & 0.8 & 0.9 & 1.0 & 1.1 \\
\hline
$-\Delta$ & 22 & 23 & 29 & 22 \\
\hline 
$-V$ & 1.4 & 1.7 & 1.8 & 1.7 \\
\hline 
\end{tabular}\\
\hline
\hline 
15 & 
\begin{tabular}{|r|llll|}
\hline
$\ve_{F}$ & 0.8 & 0.9 & 1.0 & 1.1 \\
\hline
$-\Delta$ & 62 & 62 & 61 & 61 \\
\hline 
$-V$ & 2.7 & 2.7 & 2.7 & 2.9 \\
\hline 
\end{tabular}
& 
\begin{tabular}{|r|llll|}
\hline
$\ve_{F}$ & 0.8 & 0.9 & 1.0 & 1.1 \\
\hline
$-\Delta$ & 29 & 28 & 35 & 27  \\
\hline 
$-V$ & 2.1 & 2.1 & 2.5 & 1.2 \\
\hline 
\end{tabular}
&
\begin{tabular}{|r|llll|}
\hline
$\ve_{F}$ & 0.8 & 0.9 & 1.0 & 1.1 \\
\hline
$-\Delta$ & 16 & 19 & 21 & 21 \\
\hline 
$-V$ & 1.7 & 1.6 & 1.8 & 2.0 \\
\hline 
\end{tabular}\\
\hline
\hline
25 & \begin{tabular}{|r|llll|}
\hline
$\ve_{F}$ & 0.8 & 0.9 & 1.0 & 1.1 \\
\hline
$-\Delta$ & 35 & 38 & 38 & 38 \\
\hline 
$-V$ & 2.3 & 2.4 & 2.4 & 2.8 \\
\hline 
\end{tabular}
& 
\begin{tabular}{|r|llll|}
\hline
$\ve_{F}$ & 0.8 & 0.9 & 1.0 & 1.1 \\
\hline
$-\Delta$ & 17 & 17 & 21 & 16 \\
\hline 
$-V$ & 1.7 & 1.7 & 2.0 & 1.8 \\
\hline 
\end{tabular}
&
\begin{tabular}{|r|llll|}
\hline
$\ve_{F}$ & 0.8 & 0.9 & 1.0 & 1.1 \\
\hline
$-\Delta$ & 12 & 10 & 15 & 11 \\
\hline 
$-V$ & 1.2 & 1.7 & 1.8 & 1.4 \\
\hline 
\end{tabular}\\
\hline 
\end{tabular}
\end{center} 
Table I. {\footnotesize Pair binding energy $-\D$ (in units of 
$10^{-3}t$) and Average Effective 
Interaction $V$ (in units of $t$) for $(N,N)$ nanotubes of various lengths $L$, as a 
function of the Fermi energy $\ve_{F}$ (in units of $t$). Numerical values were computed 
$U/t=2.5$ for illustration.}
\bigskip

We considered supercells of $2 \, N \times L= N_{C} $ cells, where $L$ 
is the length of the $(N,N)$ nanotube in units of the lattice spacing. 
We solved the Cooper-like equation in a virtually exact way 
for $N$ up to 6 and $L$ up to 25, using $U/t=2.5$ 
(which is of the correct order of magnitude for  
graphite\cite{lopez}\cite{hoffman}). The canonical transformation 
overestimates $\D$ in this range of $U/t$, but remains qualitatively 
correct.

The calculations are performed with 
the Fermi energy $\ve_{F}$ varying between 0.8 $t$ and 1.1 $t$ (half filling 
corresponds to $\ve_{F}=0$). As in the  $(1,1)$ cluster, the $W=0$  singlets 
show pairing (see Table I) albeit in general in $^{1}A_{2}$, as 
expected. The binding energy 
$-\Delta$ of the pairs decreases monotonically both with the radius and the length 
of the tube.  
\begin{figure}[H]
\begin{center}	
	\epsfig{figure=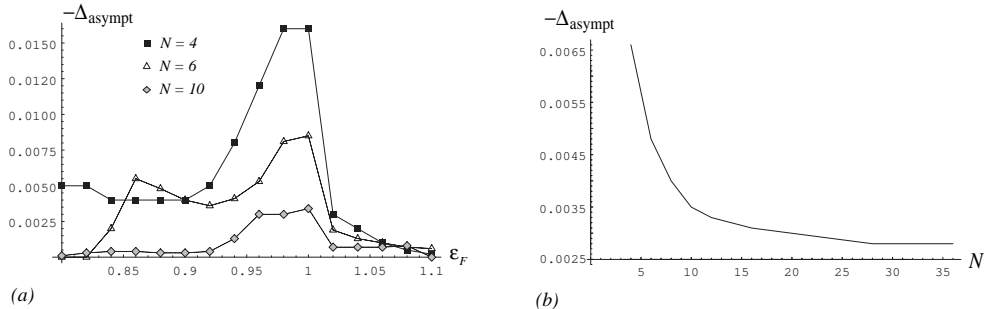,width=13cm}\caption{\footnotesize{($a$) 
	Results of the canonical transformation approach with $U/t=2.5$. 
	 $-\Delta_{\rm asympt}$ as a function of the Fermi energy $\ve_{F}$
	for $N=4$ (black boxes), $N=6$ (empty triangles) and $N=10$ (grey 
	diamonds). The Fermi energy varies in the range $0.8\div 1.1$ $t$. 
	($b$) $-\Delta_{\rm asympt}$ as a function of $N$ for $N$ 
	in the range 6$\div$36 with $\ve_{F}=t$ and Average Effective 
	Interaction $V=1.5\;t$. In both figures $-\Delta_{\rm 
	asympt}$ is in units of $t$.
	}}
\label{deltaenef}
\end{center}
\end{figure}

With supercell sizes $N_{C}>300$ numerical calculations become hard. 
Since we are concerned with the asymptotic behaviour for fixed $N$ and 
$L \to \infty $ and $\Delta(N,L)$ depends on $N$ and $L$ in a complicated way, 
we need a method to make reliable extrapolations of the numerical results. 
To this end, like in previous work\cite{EPJB1999},\cite{fettes} we define the 
Average Effective Interaction $V$. This is  such that setting in Eq.(\ref{cooplike}) 
$W_{\rm eff}= -\frac{V}{N_{C}}$, with a constant $V>0$ for all ${\bf k}$ and ${\bf k}'$ 
in ${\cal D}/4$, one obtains the correct value of $\D$. In other terms, once the 
binding energy $-\D(N,L)$ is known, the constant $V$ must be chosen in such a way that 
\begin{equation}
\frac{1}{V}=\frac{1}{N_{C}}\sum_{{\bf k}\in {\cal D}/4}\,
\frac{1}{[2\varepsilon({\bf k})+F({\bf k})]-[2\ve_{F}+F_{\rm min}({\bf 
k}_{F})]-\D(N,L)}.
\label{uim}
\end{equation}

In Table I we have reported  $V$ values;  these remain fairly stable
around $\approx 1.5\div 2$ $t$ for $N>2$  with increasing  $L$. Therefore $V$ 
is largely independent on the Fermi energy and on the radius and this allows us to 
extrapolate to  $\Delta_{\rm asympt}(N)=\lim_{L \rightarrow \infty}\D(N,L)$. 
For $N=4$ and $N=6$ we use for the average effective interaction $V$ 
the arithmetical mean of the $V$ values reported in Table I for 
$L=10,\;15$ and 25; the results are shown in Fig.(\ref{deltaenef}.$a$) 
together with $\Delta_{\rm asympt}(10)$ computed with $V=1.5\;t$. 
We found that $\Delta_{\rm asympt}$ is strongly dependent on the filling at fixed 
$N$; the sharp maximum at the {\em optimal doping} $\ve_{F}\approx 
t$ (which corresponds to a number of electrons per graphite atom of 
1.25) can be understood in terms of a corresponding peak in the density 
of states. In the {\em optimally doped} case $-\Delta_{\rm asympt}(N)$ 
decreases monotonically as the radius of the tube increases, see Fig.(\ref{deltaenef}.$b$). 
The decreasing of the binding energy 
with $N$ is suggested by recent measurements on nanotubes with diameter of 
few Angstrom\cite{tang}. However, in the limit of large $N$, 
$\Delta_{\rm asympt}(N)$ remains stable around 0.0028 $t$ and 
may be interpreted as the binding energy of the $W=0$ pair in 
an {\em optimally doped} graphite sheet. 

The paired state we have obtained here is essentially two-dimensional, 
that is the transverse direction is crucial to have a non-Abelian symmetry 
group and hence $W=0$ pairs; the pairing mechanism uses degenerate electronic 
states that exist in 2$d$ and works  away from half filling. This opens up the interesting 
possibility that in nanotubes two distinct superconducting order parameters appear in the phase 
diagram, if it turns out that  close to half-filling there is another one due to a 
breakdown of the Luttinger liquid\cite{krotov}.

Currently, intercalated graphite and carbon nanotubes superconduct at much
lower temperatures than high-$T_{c}$ Cuprates and the two kinds of materials 
are apparently quite different. However, symmetry arguments based on  the $W=0$ 
theorem tell us that, despite the obvious differences, part of the story must be 
the same, i.e. by a suitable choice of Dirac's characters the on-site 
Coulomb interaction is utterly turned off.  This produces the singlet 
pairing and constrains the ground state spin-orbital symmetry of the 
interacting system. 

}

\end{document}